\documentclass[showpacs,preprintnumbers,amsmath,amssymb,twocolumn,nobibnotes,longbibliography]{revtex4-2}%\documentclass[12pt,nonbibnotes,nofootinbib,notitlepage,longbibliography]{revtex4-1}
\usepackage{amsmath}
\usepackage{amssymb}
\usepackage{bm}
\usepackage{epsfig}
\usepackage{graphicx}
\usepackage{color}
\usepackage[unicode=true,colorlinks=true,citecolor=blue]{hyperref}
\usepackage[T2A]{fontenc}
\usepackage[cp1251]{inputenc}
\usepackage[english]{babel}
\usepackage{physics}

\usepackage[bbgreekl]{mathbbol}

\usepackage[normalem]{ulem}

\begin{document}

\title{Raman Photogalvanic Effect: photocurrent at inelastic light scattering}

\author{L. E. Golub and M. M. Glazov}
\affiliation{Ioffe Institute, 194021 St.~Petersburg, Russia}

%\date{\today}

\begin{abstract}
We show theoretically that electromagnetic waves propagating in the transparency region of a non-centrosymmetric medium can induce a \emph{dc} electric current. The origin of the effect is the Raman scattering of light by free carriers in the system. Due to the photon scattering, electrons undergo real quantum transitions resulting in the formation of their anisotropic momentum distribution and in shifts of electronic wavepackets giving rise to a steady state photocurrent. We present microscopic theory of the Raman Photogalvanic effect (RPGE) focusing on two specific situations: (i) generic case of a bulk gyrotropic semiconductor and (ii) a quantum well structure where the light is scattered by intersubband excitations. We uncover the relation of the predicted RPGE and the traditional photogalvanic effect at the light absorption.
\end{abstract}

\maketitle

\emph{Introduction.} Photogalvanic and photon drag effects {(PGE and PDE)} resulting in the \emph{dc} electric current generation under steady-state illumination belong to a class of non-linear high-frequency transport phenomena and bridge optics and transport~\cite{doi:10.1063/1.1655453,IP_CPGE,BELINICHER1978213,sturmanBOOK,ivchenko05a,reviewG,dyakonov_book,Glazov2014101}. The \emph{dc} current magnitude and direction depend on the light intensity, propagation direction, and polarization. The processes of the photocurrent generation are highly sensitive to the symmetry of the system, fine structure of the electron energy spectrum, and microscopic processes of the optical transitions and scattering~\cite{sturmanBOOK,ivchenko05a}. Furthermore, the photocurrent generation can be related in some cases to the topological properties of the charge carriers Bloch functions~\cite{Moore_2010,PhysRevB.94.035117,Juan:2017tm,LG_EL_Spivak_PRB2020,Annurev2021}. It makes polarization-dependent photocurrents an important tool to study the delicate features of the electronic spectrum and kinetic properties of charge carriers in metals and semiconductors and opens up prospects to develop polarization detectors based on these effects~\cite{doi:10.1063/1.2937192,doi:10.1063/1.3056393}.

\begin{figure}[t]
\includegraphics[width=\linewidth]{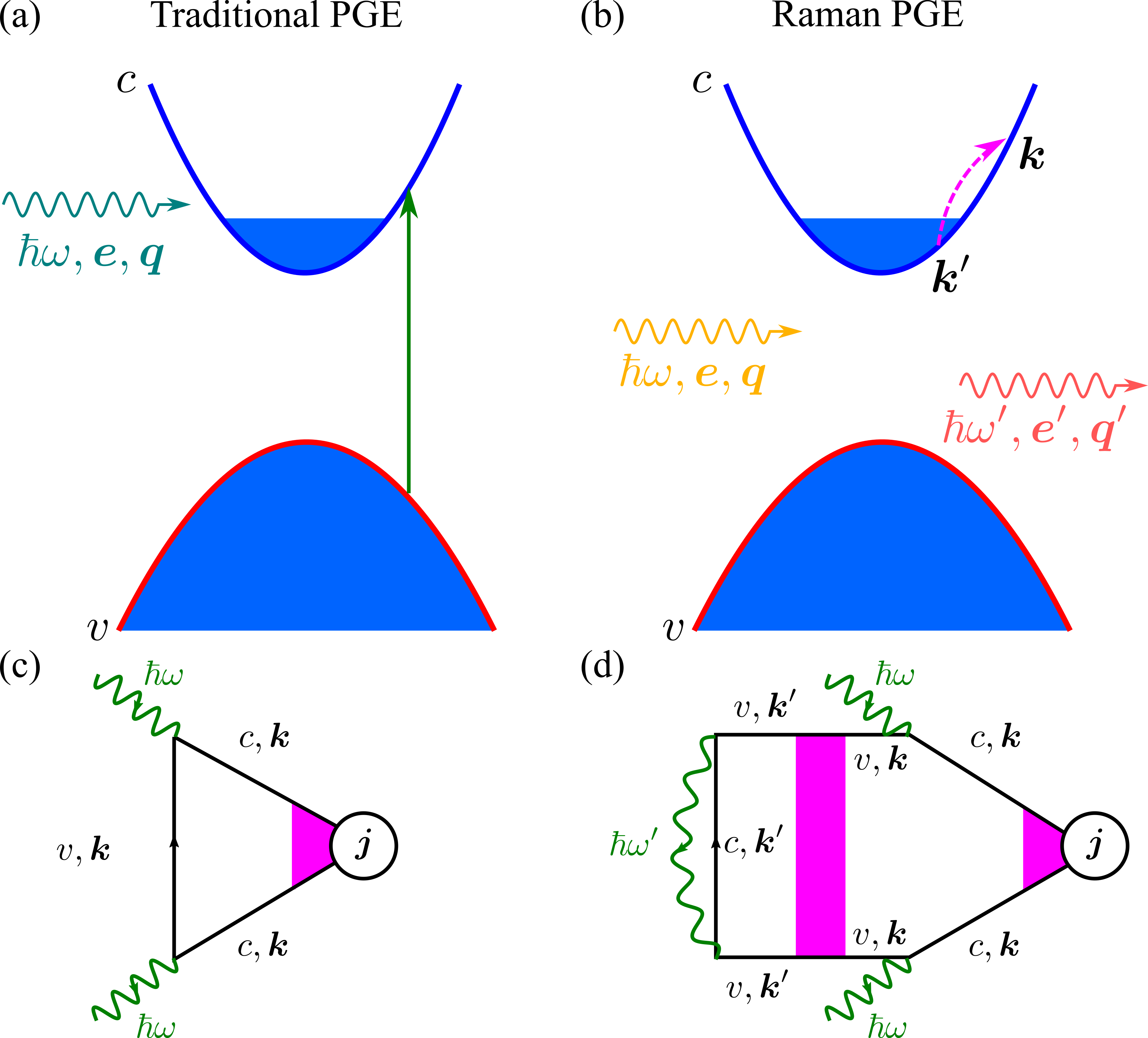}
\caption{(a) Scheme of the absorption process leading to circular PGE.
(b) Illustration of the scattering process resulting in real electronic transition. (c,d) Basic diagrams describing the effects depicted in panels above (see SI for details). Green wavy lines denote electromagnetic field, \textcircled{$\bm j$} denotes the current vertex with magenta filling describing the sum of the ladder diagrams accounting for the electron scattering by impurities and {phonons}. Magenta rectangle denotes the sum of the ladder diagrams accounting for the valence band hole scattering.}\label{fig:intro}
\end{figure}

Usually, photogalvanic and photon drag currents are observed under the conditions of light absorption, see Fig.~\ref{fig:intro}(a). The PGEs are studied in conventional bulk semiconductors like Te and GaAs~\cite{asnin_cpge,sturmanBOOK}, in low-dimensional structures such as quantum wells~\cite{PhysRevLett.86.4358,GANICHEV2002166}, and in a wide range of emergent material systems including topological insulators~\cite{Wittmann_2010,McIver2012,PhysRevLett.113.096601,PDE_TI_2015,durnev2019high,Leppenen2022}, Weyl semimetals~\cite{Ma:2017ux,Quant_CPGE_WSM}, 
graphene based nanosystems~\cite{DrexlerC.:2013uq,Edge_PGE_graphene,Edge_PGE_graphene_nonlinear,Glazov2014101},
and transition metal dichalcogenides~\cite{Quereda:2018vj,Rasmita:20,Culcer_strained_TMD_2022}. Both inter- and intraband optical transitions can be involved in the \emph{dc} current generation.

It is commonly assumed that, if the light propagates in the transparency region of the crystal, no \emph{dc} current is formed~\cite{ISI:A1986E501100021,sturmanBOOK,LG_EL_transient}. It is indeed the case provided real electronic transitions {and corresponding changes of the electromagnetic field} are absent in the system. In such a situation, the irradiation results solely in renormalization of the energy dispersion. Thus, after a transient process the current vanishes~\cite{ISI:A1986E501100021}; otherwise, in violation of the energy conservation law, such current could generate the Joule heating in the external circuit.

Here we show that even in the absence of photon absorption, the \emph{dc} electric current can be generated if the light is scattered by the free carriers in the medium. The Raman scattering of light leads to the electronic transitions, Fig.~\ref{fig:intro}(b), resulting in the asymmetry of the electron distribution in the steady state and, eventually, in the \emph{dc} current. A similar idea has been put forward in Ref.~\cite{Onishi2022} without detailed analysis, here we present an explanation of the effect and transparent microscopic model.
We develop the microscopic theory of the Raman photogalvanic effect (RPGE) in noncentrosymmetric semiconductors and semiconductor nanostructures. We mainly focus on the case of the circular RPGE where the current reverses its sign under reversal of the radiation helicity. We address the situation where the photon energy is smaller than the fundamental energy gap. We take as examples (i) a non-resonant Raman scattering in bulk semiconductors and (ii) the intersubband resonant Raman scattering in quantum wells.

\emph{General description.} We recall that the \emph{dc} current linear in the radiation intensty $I$ arising in the non-centrosymmetric media can be written in the most general form as~\cite{sturmanBOOK,ivchenko05a}
\begin{equation}
\label{PGE}
j_\alpha = \gamma_{\alpha\beta} \mathrm i [\bm e\times \bm e^*]_\beta I + \chi_{\alpha\beta\mu} (e_\beta e_\mu^* + e_\mu e_\beta^*) I,
\end{equation}
where $\bm e$ is the complex polarization vector of the incident electromagnetic field, $\mathrm i [\bm e\times \bm e^*] = P_{\rm circ} {\hat{\bm n}}$ with $P_{\rm circ}$ being the circular polarization degree and ${\hat{\bm n}}$ being the unit vector along the light propagation axis describes the light helicity. Tensors $\gamma_{\alpha\beta}$ and $\chi_{\alpha\beta\mu} = \chi_{\alpha\mu\beta}$ describe circular and linear photocurrents,
respectively; $\alpha,\beta,\mu$ are the Cartesian components. Notably, the first and second terms in Eq.~\eqref{PGE} have different properties under time reversal, $t\to-t$. Particularly, since both current and light helicity change their signs at the time reversal, tensor $\gamma_{\alpha\beta}$ is even and the tensor $\chi_{\alpha\beta\mu}$ is odd at $t\to -t$.

We assume that the light propagates in the transparency region of the direct-gap semiconductor
\begin{equation}
\label{transp}
\hbar\omega < E_g, \quad \omega \tau_p \gg 1,
\end{equation}
where $\omega$ is the frequency of radiation, $E_g$ is the band gap, and $\tau_p$ is the conduction electron momentum scattering time. Here, for definiteness, we assume that the system is $n$-doped. First condition in Eq.~\eqref{transp} ensures that the real interband transitions are forbidden, while the second one allows us to neglect the intraband Drude-like absorption. 

Under condition~\eqref{transp} absorption of light is absent and the only possible real processes are the free-carrier light scattering as illustrated in Fig.~\ref{fig:intro}(a). The incident photon with the frequency $\hbar\omega$, polarization $\bm e$, and wavevector $\bm q$ scatters and gives rise to a secondary photon with the frequency $\hbar\omega'$, polarization $\bm e'$ and wavevector $\bm q'$ while a resident electron undergoes a transition from the $\bm k'$ to $\bm k$ state. Thus, the \emph{dc} current density can be readily expressed as
\begin{equation}
\label{dc}
\bm j = \frac{e}{\mathcal V_0} \Tr\{\hat{\bm v} \rho^{(2)} \} = \bm j_b + \bm j_s,
\end{equation}
with the ballistic
\begin{subequations}
\label{components}
\begin{multline}
\label{ballistic}
\bm j_b = {\frac{2e}{\mathcal V_0}}\sum_{\bm k,\bm k', {\bm q'}} [\bm v_{\bm k} \tau_{p}(E_{c,\bm k}) - \bm v_{\bm k'} \tau_{p}(E_{c,\bm k'})] \\
\times W^{sc}_{\bm k,\bm k'}(\bm e) {f_0(E_{c,\bm k'})[1 - f_0(E_{c,\bm k})]},
\end{multline}
and shift 
\begin{multline}
\label{shift}
\bm j_s =  {\frac{2e}{\mathcal V_0}}\sum_{\bm k,\bm k', {\bm q'}} \bm R_{\bm k,\bm k'} 
\\ \times 
W^{sc}_{\bm k,\bm k'}(\bm e) {f_0(E_{c,\bm k'})[1 - f_0(E_{c,\bm k})]},
\end{multline}
\end{subequations}
contributions, respectively~\cite{sturmanBOOK,ivchenko05a,belinicher82,Sturman_2020}. Here $e$ is the electron charge, $\mathcal V_0$ is the normalization volume, $\hat{\bm v}$ is the velocity operator and $\rho^{(2)}$ is the electron density matrix calculated in the second-order in the incident electromagnetic field amplitude. In Eqs.~\eqref{components}, factors `2' account for the spin degeneracy, $E_{c,\bm k}$ is the electron dispersion, $f_0(E)$ is the equilibrium Fermi-Dirac distribution function,  $\bm v_{\bm k} = \hbar^{-1} \partial E_{c,\bm k}/\partial \bm k$ is the electron velocity, $W^{sc}_{\bm k,\bm k'}(\bm e)$ is the probability of electron scattering $\bm k'\to \bm k$ at the incident light polarization $\bm e$ averaged over the polarization and propagation direction of the final photon, and $\bm R_{\bm k,\bm k'}$ is the electron shift at the quantum transition $\bm k'\to \bm k$. 

The lack of the inversion center allows for odd in the electron wavevector terms in $W^{sc}_{\bm k,\bm k'}(\bm e)$ and even in the wavevector terms in the $\bm R_{\bm k,\bm k'}$. It makes contributions~\eqref{components} non-zero. The contribution~\eqref{ballistic} has a clear physical interpretation: In the course of quantum transitions electrons acquire an ``average'' velocity $\bar{\mathbf{v}}$ depending on the light polarization. The velocity generation rate is given by the rate of electron transitions $\dot N$ and the velocity relaxation rate is given by the momentum relaxation rate $\tau_p^{-1}$. As a result, the \emph{dc} current according to this mechanism is formed during the ballistic propagation of electrons between {the} scattering events and given by the balance of the generation and relaxation processes~\cite{sturmanBOOK}
\begin{equation}
\label{j:qual}
\bm j = e \bar{\mathbf{v}} \tau_p \dot N.
\end{equation}
The shift photocurrent in Eq.~\eqref{shift} can be estimated in the same manner with the replacement in Eq.~\eqref{j:qual} $\bar{\mathbf{v}} \tau_p$ by the ``average'' shift of the wavepacket $\bar{\mathbf{R}}$ in the course of scattering~\cite{sturmanBOOK}. Equation~\eqref{j:qual} underlies that the current generation requires real electronic transitions and $\dot N$ can be expressed via the light intensity and the extinction coefficient $\mathcal K$ related to the scattering process $\dot N = \mathcal K I/\hbar\omega$. 
%Now we turn to the microscopic description of the scattering processes.

\begin{figure}[h]
\includegraphics[width=\linewidth]{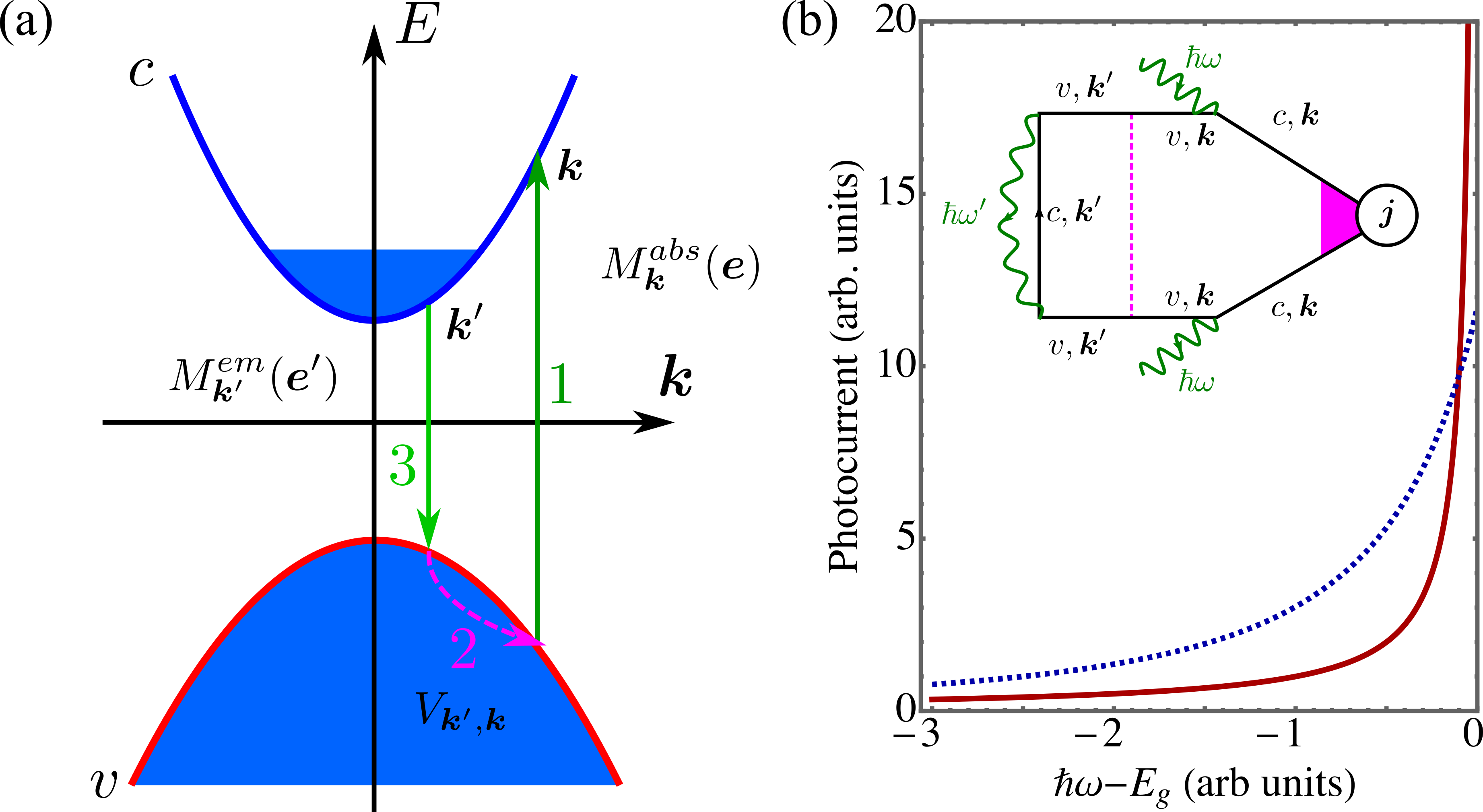}
\caption{(a)  Scheme of the light scattering process with allowance for the momentum relaxation. (b) Spectral dependence of the photocurrent for a bulk semiconductor (solid line) and quantum wells (dotted line). Inset shows the relevant diagram, see SI. }\label{fig:scheme}
\end{figure}

\emph{Microscopic model of the impurity or phonon-assisted RPGE.} Now we turn to the microscopic description of the scattering processes.
We focus on the photocurrent generation process in the bulk gyrotropic semiconductor under assumption that the incident photon energy is smaller but close to the fundamental bandgap $\Delta = E_g - \hbar\omega \ll \hbar\omega$. In this situation the main contribution to the free-carrier scattering of light is provided by virtual states in the valence band~\cite{ivchenko05a}. Accordingly, the scattering can be described as a three-stage process, where as shown in Fig.~\ref{fig:scheme}(a) (i) the incident photon is absorbed (virtually) and creates an electron-hole pair by promoting the electron with the wavevector $\bm k$ from the valence band to the conduction band; (ii) the hole in the valence band scatters (by phonon or impurity) in such a way that the state $\bm k$ in the valence band becomes filled with electron and the state with the wavevector $\bm k'$ becomes unoccupied; and (iii) the hole recombines with the resident electron, so that finally the valence band remains unperturbed (all the states are filled) and in the conduction band the state with the wavevector $\bm k'$ is empty and the state with the wavevector $\bm k$ is filled.
The corresponding scattering rate is given by 
\begin{multline}
\label{W:sc:gen}
W^{sc}_{\bm k,\bm k'}(\bm e', \bm e)= \frac{2\pi}{\hbar}\delta(\hbar\omega - \hbar\omega' - E_{c,\bm k} + E_{c,\bm k'}) I
 \\ \times\left| \frac{M^{em}_{\bm k'}(\bm e') V_{\bm k, \bm k'} M^{abs}_{\bm k}(\bm e)}{(\hbar\omega - E_{c,\bm k} + E_{v,\bm k})({\hbar\omega - E_{c,\bm k} + E_{v,\bm k'}})}\right|^2,
\end{multline}
where $M^{abs}_{\bm k}(\bm e)\sqrt{I}$ and $M^{em}_{\bm k'}(\bm e')$ are the interband transition matrix elements describing the absorption and emission of photons, respectively. It is convenient to present 
\begin{equation}
\label{Mabs:2}
\left|M^{abs}_{\bm k}(\bm e)\right|^2 = |M_0|^2(1+ D_{\alpha\beta} k_\alpha \mathrm i [\bm e\times \bm e^*]_\beta),
\end{equation}
where the real second-rank tensor $D_{\alpha\beta}$ is responsible for the gyrotropy of the system~\cite{sturmanBOOK} and $M_0$ is a constant. The presence of $\bm k$-linear terms in Eq.~\eqref{Mabs:2} makes an odd in the wavevector contribution to the electron transition rate $W^{sc}_{\bm k,\bm k'}(\bm e) = \sum_{\bm q', \bm e'} W^{sc}_{\bm k,\bm k'}(\bm e',\bm e)$ and eventually results in the non-zero photocurrent, Eq.~\eqref{shift}. Equations~\eqref{ballistic} and \eqref{W:sc:gen} correspond to the diagram in Fig.~\ref{fig:scheme}(b). Assuming that the carrier's momentum relaxation is caused by short-range impurities, introducing $\xi$ as the ratio of the conduction and valence band elastic scattering matrix elements squared and $\gamma_r$  as the radiative decay rate of the photoexcited electron-hole pair we obtain the following expression for the circular photocurrent (see SI for details of derivation)
\begin{equation}
\label{j:ball:compact}
j_\alpha = e n_e \gamma_r D_{\alpha\beta} {\hat n_\beta P_{\rm circ}} I 
%\frac{|M_0|^2 \xi}{3\pi\Delta^2}\Phi(\nu)
\abs{{M_0}\over\Delta}^2{\xi \Phi(\nu) \over 3\pi}.
\end{equation}
Here $n_e$ is the electron density in the conduction band, $\nu = 1+ m_e/m_h$ with $m_e$ and $m_h$ being the electron and hole effective masses, and $\Phi(\nu)= [(\nu+1)\ln\nu - 2(\nu-1)](\nu-1)^{-3}$. In derivation of Eq.~\eqref{j:ball:compact} we assumed degenerate electrons with their Fermi energy $E_{\rm F} \ll \Delta$. The photocurrent in Eq.~\eqref{j:ball:compact} increases with decreasing the detuning $\Delta$ because the smaller $\Delta$ is the more efficient is the light scattering, see Fig.~\ref{fig:scheme}(b).

The ballistic photocurrent~\eqref{j:ball:compact} changes its sign at reversal of the radiation helicity and Eq.~\eqref{j:ball:compact} describes the circular RPGE. To obtain the ballistic linear RPGE one has to go beyond the three-stage process described above and take into account additional scattering processes to ensure the correct properties of the current under time reversal or evaluate the shift contribution, Eq.~\eqref{shift}, see Ref.~\cite{golub2011shift} for evaluation of $\bm R_{\bm k,\bm k'}$ for multiquantum transitions. In any case, the  linear RPGE will have an additional smallness $\sim (\Delta \tau_p/\hbar)^{-1}$, $(\omega\tau_p)^{-1}$, $(E_{\rm F} \tau_p/\hbar)^{-1}$ depending on the particular mechanism of the effect.

It is worth to mention that the main contribution to the Raman scattering of light by free charge carriers in semiconductors does not require an additional transition of the hole in the valence band. The electron wavevector can change due to the variation of the light wavevector in the course of scattering
\begin{equation}
\label{momentum}
\bm k - \bm k' = \bm q - \bm q',
\end{equation}
see SI for details. This process is described by the diagram analogous to that in Fig.~\ref{fig:scheme}(b) but without the dashed vertical line. Calculation shows that the resulting photocurrent differs from that derived above in Eq.~\eqref{j:ball:compact} by a factor of $\hbar q^2 \tau_p/m_e\ll 1$. The smallness of such contribution is related to the fact that, in the absence of additional scattering of the hole, the initial and final wavevectors of electrons are close to each other, see Eq.~\eqref{momentum}: The electron wavevector cannot change more than by a radiation wavevector. It results in a significant reduction of the effect.

\begin{figure}[h]
\includegraphics[width=0.99\linewidth]{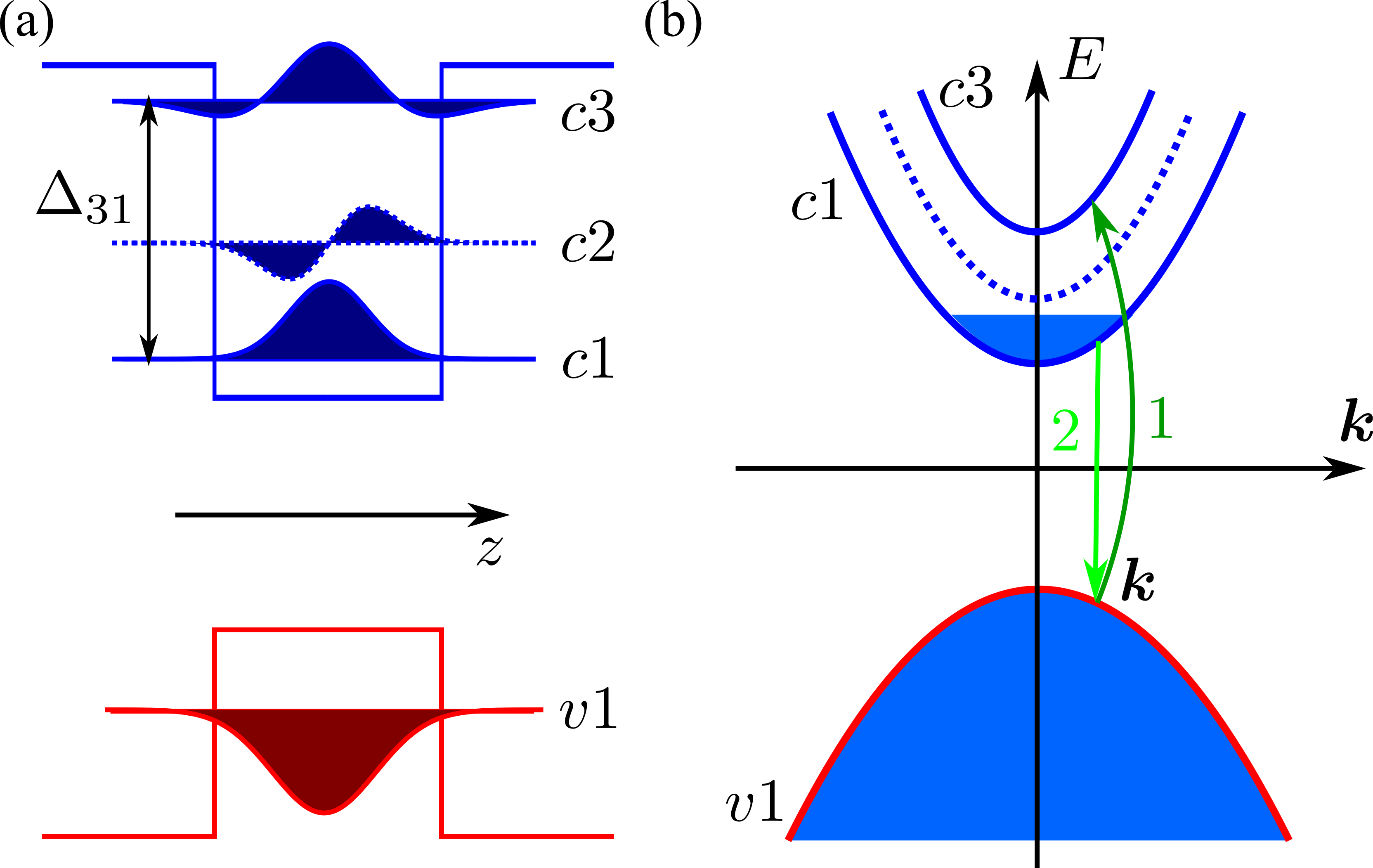}
\caption{(a) Quantum well structure. (b) Scheme of the intersubband resonant light scattering process.}\label{fig:QW:merged}
\end{figure}

\emph{RPGE at the intersubband resonance in quantum well structures.} Additional specifics of Raman scattering of light appears where the change of the photon frequency corresponds to a frequency of a resonant excitation in the system. Such situation can naturally arise in quantum well structures as shown in Fig.~\ref{fig:QW:merged} under conditions of the intersubband scattering~\cite{ivchenko05a}. We consider for simplicity a symmetric structure with lowest occupied conduction subband ($c1$). As before the frequency of incident photon corresponds to the transparency region of the structure, Eq.~\eqref{transp} and $E_g$ in this case corresponds to the gap between the topmost valence subband $v1$ and the bottom conduction subband $c1$. A situation of interest occurs in the vicinity of the intersubband resonance
\begin{equation}
\label{energ}
\hbar\omega - \hbar\omega' = E_{c3,\bm k} - E_{c1,\bm k},
\end{equation}
where the process depicted in Fig.~\ref{fig:QW:merged}(b) becomes possible. In this case the virtual photon absorption via $v1\to c3$ electron transition is followed by the photon emission process resulting from the $c1\to v1$ transition. As a result, an electron is promoted from $c1\to c3$ subband (in asymmetric quantum wells similar transitions involving $c2$ subband are also allowed). Corresponding transition rate is readily evaluated [cf. Eq.~\eqref{W:sc:gen}]
\begin{multline}
\label{intersubband}
W^{sc}_{3\leftarrow 1}(\bm e', \bm e, \bm k)
 = \frac{2\pi}{\hbar}\left| \frac{M^{em}_{\bm k}(\bm e')  M^{abs}_{\bm k}(\bm e)}{\hbar\omega - E_{c3,\bm k} + E_{v1,\bm k}}\right|^2\\
 \times \delta(\hbar\omega - \hbar\omega' - E_{c3,\bm k} + E_{c1,\bm k}) I.
\end{multline}
We neglect the photon momentum and any additional phonon or impurity scattering processes, hence the transitions take place at the same electron wavevector $\bm k$. As above we focus on the ballistic circular RPGE because it dominates the photocurrent and using the same Eq.~\eqref{Mabs:2} for $|M^{abs}_{\bm k}(\bm e)|^2$  as in the bulk case we arrive at the following expression for the current density (see SI for details):
\begin{equation}
\label{j:inter}
j_\alpha = - e n_e \gamma_r^{\rm QW} D_{\alpha\beta} {\hat n_\beta P_{\rm circ}}   I {|M_0|^2 \tau_\text{tr} \over  \nu_{31}\Delta_{QW} \hbar }  \chi(\varepsilon).
\end{equation}
Here $\gamma_r^{\rm QW}$ is the recombination rate of the electron in $c1$ subband with the $v1$ hole,  we assumed the parabolic dispersion $E_{c1,\bm k} = \hbar^2 k^2/2m_{c1}$, $E_{c3,\bm k} = \Delta_{31} + \hbar^2 k^2/2m_{c3}$ with $m_{ci}$ ($i=1,3$) being the effective mass in the $i$th subband, and $\Delta_{31}$ being the intersubband energy gap. In Eq.~\eqref{j:inter} we have introduced $\nu_{31}= m_{c1}/m_{c3}+ m_{c1}/m_h$,  $\tau_{\rm tr}$ is the momentum scattering time of $c1$ electron at a Fermi surface, the detuning for quantum wells $\Delta_{QW}=E_{g} + \Delta_{31} -\hbar\omega$, and
\begin{equation}
\varepsilon = {\Delta_{QW}\over \nu_{31}E_\text{F}}, \quad \chi(\varepsilon) = \varepsilon\qty(\ln{1+\varepsilon\over \varepsilon}-{1\over 1+\varepsilon}).
\end{equation}
Equation~\eqref{j:inter} is valid for degenerate electrons and under assumption that the momentum relaxation in the $c3$ subband is much faster than in $c1$ subband: It is typically the case because of optical phonon emission processes causing electrons to relax to the bottom subbands; the general case is considered in the SI.
It follows from Eq.~\eqref{j:inter} that RPGE current at intersubband scattering tends to a constant at the absorption edge being much weaker function of the detuning as compared to the analogous photocurrent in the bulk, cf. Eq.~\eqref{j:ball:compact} and Fig.~\ref{fig:scheme}.

\emph{Comparison with circular PGE in the absorption region.} It is instructive to compare the results for the circular photocurrent obtained here for the transparency region with the well-known results for the circular PGE at the direct interband transitions. Considering the bulk semiconductor in the model described above [see diagram in Fig.~\ref{fig:intro}(c)] we obtain the following expression for the conduction electron photocurrent at $\hbar\omega> E_g$:
\begin{equation}
\label{CPGE:fin:abs}
j_\alpha^{\rm abs}= e {\mathcal A I\over \hbar\omega}   D_{\alpha\beta}  {\hat n_\beta P_{\rm circ}}  \frac{2|\Delta| \tau_p}{3 \nu^2\hbar} .
\end{equation}
Here $\mathcal A$ is the absorption coefficient of the semiconductor, the electron momentum scattering time $\tau_p$ is taken at the energy $|\Delta|/{\nu}$, and $\Delta = E_g -\hbar\omega<0$ in the case of direct optical transitions. One can recast Eq.~\eqref{j:ball:compact} in a similar form via the extinction coefficient $\mathcal K$
\begin{equation}
\label{j:ball:compact:K}
j_\alpha^{\rm scatt}= e {\mathcal K I\over \hbar\omega}   D_{\alpha\beta} {\hat n_\beta P_{\rm circ}} \frac{2|\Delta| \tau_p}{3 \hbar} \frac{\Phi(\nu)}{\tilde \Phi(\nu)} .
\end{equation}
with $\tilde \Phi(\nu) = \pi\sqrt{\nu}/[2(1+\sqrt{\nu})^3]$. 
The factors $\nu^2$, $\Phi/\tilde \Phi \sim 1$, as a result for the same value of detuning the traditional and Raman  photocurrents differ by the factor $\sim \mathcal A/\mathcal K$ which provides an estimate of the ratio of the electronic transition rates at the light absorption and scattering, respectively. Naturally, both expressions can be brought to the form of general Eq.~\eqref{j:qual} with 
\begin{equation}
\label{bar:v}
\bar{\rm v}_\alpha \sim  \frac{\Delta}{\hbar} D_{\alpha\beta} \hat{n}_\beta P_{\rm circ}.
\end{equation}
Equations~\eqref{j:qual}, \eqref{j:ball:compact:K}, and \eqref{bar:v} demonstrate that even for transparent media, real electronic transitions should occur to enable the photocurrent~\footnote{Further details and comparison with recent Refs. ~\cite{Onishi2022,shi2022berry} are presented in SI.}.

We also note that in gyrotropic semiconductors and nanostructures $\bm k$-linear terms are present in the effective Hamiltonian of the charge carriers due to the {spin-orbit coupling}~\cite{rashbasheka,ivchenko05a,golub_ganichev_BIASIA,dyakonov_book}:
\begin{equation}
\label{k-linear}
\mathcal H_{\rm SO} = \hbar \beta_{\alpha\mu} k_\alpha\sigma_\mu,
\end{equation}
where $\bm \sigma/2$ is the electron spin operator. These terms provide an additive mechanism for the RPGE current generation. It can be also described by the general Eq.~\eqref{j:qual} with $\bar{\rm v}_\alpha \sim \beta_{\alpha\mu} \hat{n}_{\mu} P_{\rm circ}$.

\emph{Conclusion.} We have shown that the light scattering results in the steady state current in noncentrosymmetric media. The RPGE current is generated even if the light is propagating in the transparency spectral region of the crystal. While absorption is absent in this case, the current results from real electronic transitions owing to the Raman scattering of photons by free charge carriers. These transitions cause asymmetric distribution of electrons and also quantum shifts. We have identified key mechanisms of the Raman scattering induced circular photocurrent for the photon energies slightly below the band gap of a semiconductor and studied the photocurrent generation under intersubband scattering in quantum well structures.

\emph{Acknowledgements.}
We are grateful to Y. Onishi for valuable discussions. The authors acknowledge support from RSF: Project 22-12-00211 (general theory of RPGE, M.M.G.) and  22-12-00125 (calculations of RPGE currents, L.E.G.). L.E.G. also thanks the Foundation for the Advancement of Theoretical Physics and Mathematics ``BASIS''.

\bibliography{Raman}

\end{document}

\title{
Online supplementary information:\\
Raman Photogalvanic Effect: photocurrent at inelastic light scattering
}

\author{L.~E.~Golub} 	
\author{M.~M.~Glazov}

\affiliation{Ioffe Institute,  	194021 St.~Petersburg, Russia}

\maketitle

Here we present the details of derivation of photocurrents, particularly, Eqs. (8) and (12) of the main text, present the model for the Raman photocurrent associated with the photon wavevector, which does not require electron scattering, and discuss the relation between this work and recent preprints~\cite{Onishi2022,shi2022berry}.

\tableofcontents

\section{Ballistic circular photocurrent in bulk gyrotropic semiconductor} 

%Let
%\begin{equation}
%\label{optical}
%M_{\bm k}^{abs}(\bm e) = M_0 \bm e \cdot \bm d_{\bm k}, \quad d_\alpha(\bm k) = d^0_\alpha + \mathrm i d^1_{\alpha\beta} k_\beta + \ldots,
%\end{equation}
%where $\alpha,\beta=x,y,z$ are Cartesian components, $d^0_\alpha$ describes $\bm k$-independent contribution (allowed optical transition), $d^1_{\alpha\beta}$ is responsible for gyrotropy of the system, higher order in $\bm k$ terms are denoted by \ldots. For the reasons described above we are interested only in the $\bm k$-dependence of $M_{\bm k}^{abs}(\bm e)$ and replace {$|V_{\bm k,\bm k'}|^2$ by a constant $|V|^2 n_i/\mathcal V_0$ with $n_i$ being the impurity density}. 
%As a result, we obtain ($|\bar d^0|^2$ is the polarization averaged $|\bm d^0|^2$)
%\begin{multline}
%\label{optical:1}
%W_{\bm k, \bm k'}^{sc}(\bm e) = \frac{2\pi}{\hbar} I \frac{|M_{{em}}\bar d^0 V|^2 {n_i}}{{\mathcal V_0 [\hbar\omega - E_{c}(\bm k) + E_{v}(\bm k)]^2[\hbar\omega - E_c(\bm k) + E_{v}(\bm k')]^2}} \\
%\times |M_0|^2 {\frac{1}{2}}([\bm d(\bm k) \times \bm d^*(\bm k)] \cdot[\bm e \times \bm e^*])\delta[\hbar\omega - \hbar\omega' - E_c(\bm k) + E_c(\bm k')].
%\end{multline}
%We can represent the $\bm k$-linear contribution to $([\bm d(\bm k) \times \bm d^*(\bm k)] \cdot[\bm e \times \bm e^*])$ as
%\begin{equation}
%\label{k:lin}
%{\frac{1}{2}} ([\bm d(\bm k) \times \bm d^*(\bm k)] \cdot[\bm e \times \bm e^*]) =   D_{\alpha\beta} k_\alpha \mathrm i [\bm e\times \bm e^*]_\beta,
%\end{equation}
%with the real second-rank pseudotensor $D_{\alpha\beta}$ and 

Substituting the scattering rate~(6) [of the main text] with the squared matrix element~(7) into Eq.~(4a) we
obtain the ballistic photocurrent in the form 
\begin{multline}
\label{j:ball:res}
j_\alpha =  {\frac{e}{\mathcal V_0}}|M_0|^2 |M^{em}|^2_\Sigma D_{\alpha\beta} \mathrm i [\bm e\times \bm e^*]_\beta \frac{8\pi}{3\hbar^{{2}}} I \frac{{n_i}|V|^2}{{\mathcal V_0}}\\
\times  \sum_{\bm k,\bm k',\bm q'}  {\frac{\tau_p(E_{c, \bm k}) E_{c, \bm k} f_0(E_{c, \bm k'})[1 - f_0(E_{c, \bm k})] }{(\hbar\omega - E_{c, \bm k} + E_{v, \bm k})^2(\hbar\omega - E_{c, \bm k} +E_{v, \bm k'})^2}}  \delta(\hbar\omega - \hbar\omega' - E_{c, \bm k} + E_{c, \bm k'}).
\end{multline}
Here we assumed the parabolic dispersion $E_{c, \bm k} = \hbar^2 k^2/2m_e$ with $m_e$ being the effective mass. 
Here ${|M^{em}|^2_\Sigma}$ is the emission matrix element summed over the secondary photon polarization. In calculation of ${|M^{em}|^2_\Sigma}$ we can disregard $\bm k'$-linear terms because they are sensitive to $\bm e'$, we also disregard $k'^2$ and higher-order contributions to the matrix elements. Note that $M^{abs}(\bm e)$ describes the stimulated process (we consider classical electromagnetic wave incident on the sample), while emission process is spontaneous. That is why we use different normalization of $M^{em}$ and $M^{abs}$, see the main text below Eq. (6) for details.

For the following calculations we perform the summation over $\bm q'$ by means of the energy conservation $\delta$-function:
\begin{equation}
\label{j:ball:res:1}
j_\alpha =  {\frac{e}{\mathcal V_0}}D_{\alpha\beta} \mathrm i [\bm e\times \bm e^*]_\beta \frac{4}{3{\hbar}} I \frac{n_i|M_0V|^2}{\mathcal V_0} \gamma_r  \sum_{\bm k,\bm k'}  {\frac{\tau_p(E_{c, \bm k}) E_{c, \bm k} f_0(E_{c, \bm k'})[1 - f_0(E_{c, \bm k})] }{(\hbar\omega - E_{c, \bm k} + E_{v, \bm k})^2(\hbar\omega - E_{c, \bm k} +E_{v, \bm k'})^2}},
\end{equation}
where we introduced the rate of emission of the secondary photon (i.e., the recombination rate of the electron-hole pair in the vicinity of the fundamental band gap)
\[
\gamma_r = \frac{2\pi}{\hbar} {|M^{em}|^2_\Sigma} \sum_{\bm q'} \delta(\hbar\omega - \hbar\omega' - E_{c, \bm k} + E_{c, \bm k'}).
\]
Further calculations can be simplified as follows: we sum over $\bm k'$ (initial electron wavevector) under the assumption that $E_F, T \ll E_g - \hbar\omega$ ($T$ is the temperature expressed in the energy units). Hence, the $\bm k'$-dependence of the denominators can be neglected in this case. This summation yields $n_e \mathcal V_0/2$ with 	
$n_e = (2/\mathcal V_0)\sum_{\bm k'}f_0(E_{c, \bm k'})$  being the electron density. Under the same assumption one can omit $1 - f_0(E_{c, \bm k})$ (but keep the $\bm k$-dependence in the denominators otherwise the integral diverges at large $\bm k$)
and we finally have
\begin{equation}
\label{j:ball:res:2}
j_\alpha =  {\frac{e}{\mathcal V_0}}D_{\alpha\beta} \mathrm i [\bm e\times \bm e^*]_\beta \frac{2}{3{\hbar}} I n_e {n_i|M_0V|^2} \gamma_r  \sum_{\bm k}  {\frac{\tau_p(E_{c, \bm k}) E_{c, \bm k} }{(\hbar\omega - E_{c, \bm k} + E_{v, \bm k})^2(\hbar\omega - E_{c, \bm k} +E_{v, \bm k'})^2}}.
\end{equation}
Here all energies are counted from the bottom of the conduction band.
Using
\begin{equation}
n_i\abs{V}^2\tau_p(E_c)= {\xi} {\hbar\over 2\pi g_c(E_c)}, \quad \sum_{\bm k}\ldots  = \mathcal V_0 \int_0^\infty \dd E_c g_c(E_c) \ldots, \quad
E_{c, \bm k}-E_{v, \bm k} = \nu E_{c, \bm k} + E_g, \quad \nu = 1+ {m_e\over m_h},
\end{equation}
where {$\xi$ is the ratio of the conduction and valence band elastic scattering matrix elements squared: $\xi=\abs{V_v}^2/\abs{V_c}^2$}, $g_c(E_c)$ is the density of states in the conduction band, we get
\begin{equation}
j_\alpha =  eD_{\alpha\beta} \mathrm i [\bm e\times \bm e^*]_\beta I n_e {|M_0|^2} \gamma_r {\xi\over 3\pi}\int_0^\infty \dd E_c  \frac{E_c}{(\hbar\omega - E_g - \nu E_{c})^2(\hbar\omega - E_g - E_c)^2}.
\end{equation}
Calculating the integral
\begin{multline}
\int_0^\infty \dd E_c  \frac{E_c}{(\hbar\omega - E_g - \nu E_{c})^2(\hbar\omega - E_g - E_c)^2}
= {\Phi(\nu)\over \Delta^2}, \qquad \Delta = E_g-\hbar\omega,
\\ \Phi(\nu) = \int_0^\infty \dd x  \frac{x}{(1+ \nu x)^2(1+x)^2} = {(\nu+1)\ln\nu - 2(\nu-1) \over (\nu-1)^3},
%\xrightarrow[\nu \to 1]{}  {1\over 6},
\qquad
\Phi(1) ={1\over 6},
\end{multline}
we finally get Eq.~(8) of the main text:
%\begin{equation}
%j_\alpha = eD_{\alpha\beta} \mathrm i [\bm e\times \bm e^*]_\beta \frac{2}{3{\hbar}} I n_e {|M_0|^2} \gamma_r {\xi} {\hbar\over 2\pi}{\Phi(\nu)\over (E_g-\hbar\omega)^2}.
%\end{equation}
%It is also convenient to rewrite this result as
\begin{equation}
\label{j:ball:compact}
j_\alpha = e n_e\gamma_r D_{\alpha\beta} \mathrm i [\bm e\times \bm e^*]_\beta I {|M_0|^2\over \Delta^2} \frac{\xi \Phi(\nu)}{{3\pi}}.
\end{equation}
%where we introduce the detuning $\Delta = E_g-\hbar\omega$.
% and $\Phi(\nu)= {(\nu+1)\ln\nu - 2(\nu-1) \over (\nu-1)^3}$.

\subsection{Light extinction and RPGE}

Let us introduce the extinction coefficient $\mathcal K$ [cm$^{-1}]$ related to the light scattering
\begin{equation}
\label{K:def}
\mathcal K = \frac{2\sum_{\bm k,\bm k', \bm q'} W^{sc}_{\bm k,\bm k'}(\bm e) f_0(E_{c, \bm k'})[1 - f_0(E_{c, \bm k})] }{\mathcal N (c/n)}.
\end{equation}
It describes light attenuation in the system due to the scattering. 
Here $n$ is the refractive index of the crystal and $\mathcal N$ is the number of photons in the electromagnetic wave. Taking into account that 
%$\mathcal N = |A_0|^2(\omega n/c)^2/(2\pi \hbar\omega)$ and $|A_0|^2 = 2\pi c/n\omega^2 I$ ($|A_0|$ is the amplitude of the vector potential in the incident electromagnetic wave) we have 
$I = \mathcal N \hbar\omega c/n$ and
performing summation over $\bm q'$ to obtain $\gamma_r$, over $\bm k'$ to obtain $n_e$, and over $\bm k$ by virtue of
\[
\int_0^\infty \dd E_c   \frac{g_c(E_c)}{(\hbar\omega - E_g - \nu E_{c})^2(\hbar\omega - E_g - E_c)^2} = \tilde{\Phi}(\nu)\frac{g_c(E_\omega)}{E_\omega^3 \nu^3 }, \quad E_\omega={E_g- \hbar\omega\over \nu} , \quad 
\tilde{\Phi}(\nu) = \frac{\pi\sqrt{\nu}}{2(1+\sqrt{\nu})^3},
\]
and using
$g_c({E_\omega})n_i\abs{V}^2=\xi \hbar /[ 2\pi \tau_p(E_\omega)]$,
we have
\begin{equation}
\label{K:1}
\mathcal K =
n_e \gamma_r {\xi \hbar \over 2\pi \tau_p}\tilde{\Phi}(\nu)\frac{\hbar\omega |M_0|^2}{\Delta^3} .
\end{equation}
As a result for the Raman scattering-induced photocurrent we have Eq.~(15) of the main text:
\begin{equation}
\label{j:ball:compact}
j_\alpha^{\rm scatt}= e {\mathcal K I\over \hbar\omega}   D_{\alpha\beta} \mathrm i [\bm e\times \bm e^*]_\beta \frac{2 \Delta \tau_p(E_\omega)}{3 \hbar} \frac{\Phi(\nu)}{\tilde \Phi(\nu)} .
\end{equation}

\subsection{Relation to the photogalvanic effect 
at light absorption}

Let us calculate the (ballistic) CPGE current in the same system generated at $\hbar\omega > E_g$. We calculate the electron contribution 
only (this is the net electric current if the relaxation time in the valence band is very short). The CPGE current density at light absorption is given by
\begin{equation}
\label{CPGE:0}
j_\alpha^{\rm abs} 
= 2e\sum_{\bm k} W^{abs}_{\bm k}(\bm e) \tau_p(E_{c, \bm k}) v_\alpha(\bm k) \delta(E_{c, \bm k}-E_{v, \bm k}-\hbar\omega),
\end{equation}
where the asymmetric contribution to the light absorption probability at direct optical transition is
\begin{equation}
W^{abs}_{\bm k}(\bm e) = \frac{2\pi}{\hbar} I \abs{M_0}^2  D_{\alpha\beta} \mathrm i [\bm e\times \bm e^*]_\beta k_\alpha.
\end{equation}
Then we obtain
\begin{multline}
\label{CPGE:fin}
j_\alpha^{\rm abs} = 2e \frac{2\pi}{\hbar} I \abs{M_0}^2  D_{\alpha\beta} \mathrm i [\bm e\times \bm e^*]_\beta \sum_{\bm k}  {2\over 3\hbar}E_{c, \bm k} \tau_p(E_{c, \bm k}) \delta(\nu E_{c, \bm k}+E_g-\hbar\omega)
\\
= eI   D_{\alpha\beta} \mathrm i [\bm e\times \bm e^*]_\beta \frac{8\pi\abs{M_0}^2}{3\hbar^2}  
%N_\omega  
g_c(\abs{E_\omega})
{\abs{E_\omega} \tau_p(\abs{E_\omega})\over \nu^2}  \Theta(\hbar\omega-E_g).
\end{multline}

It is convenient to introduce the light absorption coefficient $\mathcal A(\omega)$  with the dimension of cm$^{-1}$ related to the direct optical transitions as
\begin{equation}
\label{direct:A}
\mathcal A 
= \frac{\hbar\omega}{I} W_{cv},
\end{equation}
where the direct interband transition rate 
\begin{equation}
\label{direct:tr:rate}
W_{cv} = {2} \frac{2\pi}{\hbar} \sum_{\bm k} I |M_0|^2 \delta(E_{c, \bm k} - E_{v, \bm k} -\hbar\omega)  =  \frac{4\pi}{\hbar} I |M_0|^2 \frac{g_c(\abs{E_\omega})}{\nu}\Theta(\hbar\omega-E_g),
\end{equation}
and
\begin{equation}
\label{direct:A:1}
\mathcal A 
= {4\pi \omega} |M_0|^2 \frac{g_c(\abs{E_\omega})}{\nu}\Theta(\hbar\omega-E_g).
\end{equation}
Finally, we obtain from Eq.~\eqref{CPGE:fin} the circular photocurrent caused by the CPGE in the form of Eq.~(14) of the main text:
\begin{equation}
\label{CPGE:fin:abs}
j_\alpha^{\rm abs}= e {\mathcal A I \over \hbar \omega}   D_{\alpha\beta} \mathrm i [\bm e\times \bm e^*]_\beta  \frac{2 \abs{E_\omega} \tau_p(\abs{E_\omega})}{3 \nu\hbar} .
\end{equation}

\section{Allowance for the radiation wavevector in RPGE}

\begin{figure}[h]
\includegraphics[width=0.35\textwidth]{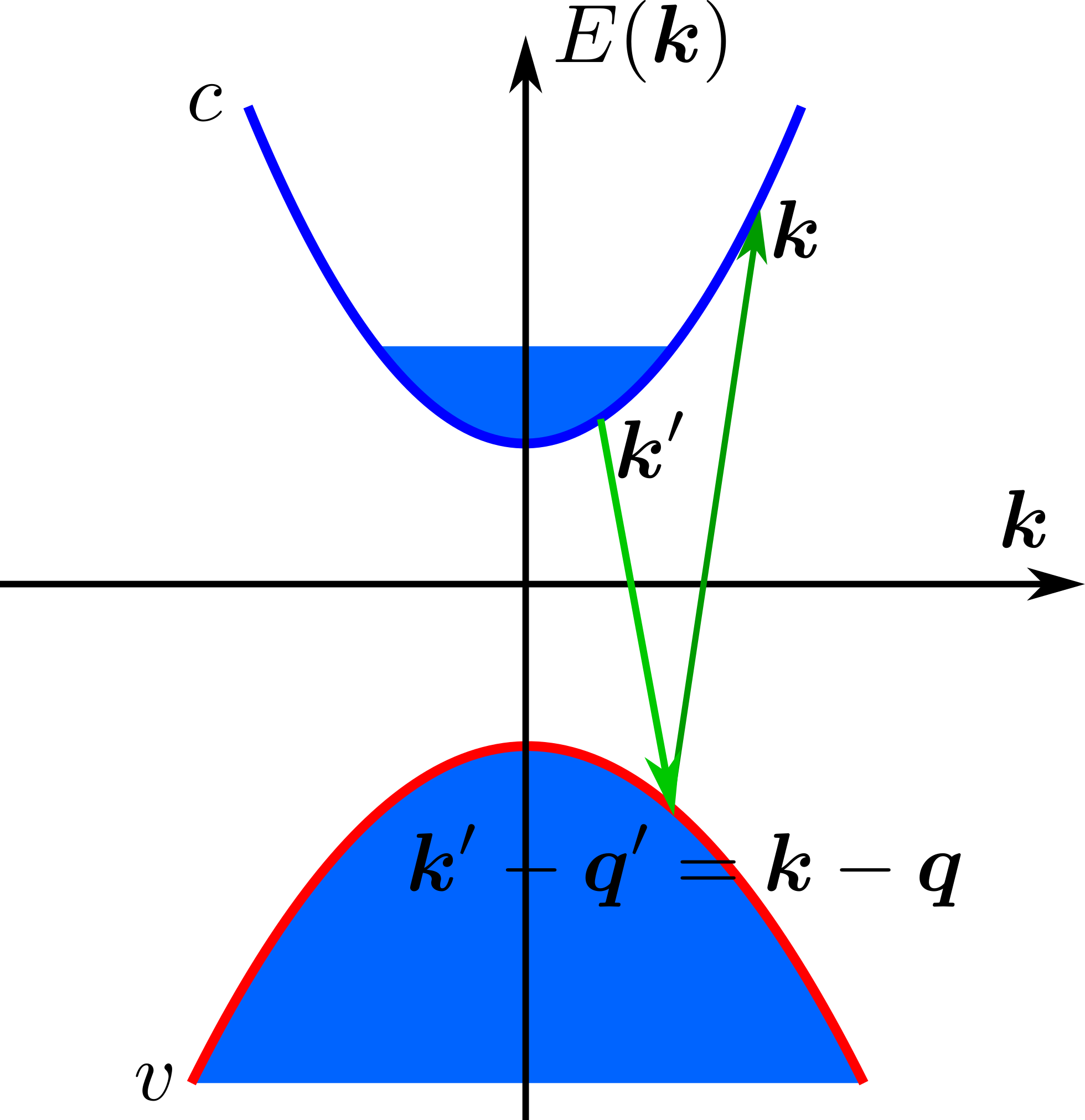}
\caption{Scheme of the light scattering process with allowance for the radiation wavevector.}\label{fig:scheme2}
\end{figure}

In previous section we studied the RPGE photocurrent that is generated in a course of a three-step process of virtual photon absorption, valence band hole scattering, and virtual photon emission. In this way the electron wavevectors in the initial, $\bm k'$, and final, $\bm k$, states are decoupled. Let us now discuss the contribution to RPGE where the valence band hole scattering is absent. Such process takes place with allowance for the photon wavevector, see Fig.~\ref{fig:scheme2}, this process is the main process of light scattering by free carriers in semiconductors for $\hbar\omega \lesssim E_g$~\cite{ivchenko05a}. Corresponding transition probability is given by
\begin{equation}
\label{W:sc:q}
W^{q}_{\bm k,\bm k'}(\bm e', \bm e)= \frac{2\pi}{\hbar}\delta(\hbar\omega - \hbar\omega' - E_{c,\bm k} + E_{c,\bm k'}) \delta_{\bm k + \bm q', \bm k' + \bm q} I \left| \frac{M^{em}_{\bm k'}(\bm e') M^{abs}_{\bm k}(\bm e)}{\hbar\omega - E_{c,\bm k} + E_{v,\bm k}}\right|^2.
\end{equation}
Here Kronecker-$\delta$ accounts for the momentum conservation in the course of light scattering [Eq. (9) of the main text]
\[
\bm k' + \bm q = \bm k + \bm q'.
\]
The photocurrent is given by [cf. Eq. (4a) of the main text]
\begin{equation}
\label{ballistic:q}
\bm j_b = {\frac{2e}{\mathcal V_0}}\sum_{\bm k,\bm k', {\bm q'}} [\bm v_{\bm k} \tau_{p}(E_{c,\bm k}) - \bm v_{\bm k'} \tau_{p}(E_{c,\bm k'})]  W^{q}_{\bm k,\bm k'}(\bm e) {f_0(E_{c,\bm k'})[1 - f_0(E_{c,\bm k})]},
\end{equation}
Let us assume for simplicity that the energy dependence of the electron momentum scattering time can be disregarded. In this case the velocity-dependent term in square brackets can be recast as
\[
[\bm v_{\bm k} \tau_{p}(E_{c,\bm k}) - \bm v_{\bm k'} \tau_{p}(E_{c,\bm k'})] = \tau_p \frac{\hbar}{m_e} (\bm q' - \bm q).
\]
Note that the difference of velocities in the initial and final states is now small and related to the photon wavevector. It results in suppression of the photocurrent.

Disregarding in Eq.~\eqref{ballistic:q} the $\bm q, \bm q'$-dependence of all remaining terms we obtain the polarization independent photocurrent related to the photon \emph{drag} effect at the Raman scattering:
\begin{equation}
\label{ballistic:drag}
\bm j_b = {\frac{2e}{\mathcal V_0}}\frac{\hbar}{m_e} (- \bm q) \sum_{\bm k, {\bm q'}}  W^{q}_{\bm k,\bm k}(\bm e) {f_0(E_{c,\bm k})[1 - f_0(E_{c,\bm k})]}.
\end{equation}
According to the time-reversal symmetry this effect is independent of the circular polarization of light.

Furthermore, to obtain the polarization-dependent current we need to extract $\bm k$- and $\bm q$-linear contributions from the occupancy factors, otherwise the sum over $\bm k$ in Eq.~\eqref{ballistic:q} vanishes. The resulting current contains the factor $\sim \hbar^2 q^2/m_e$ instead of $\sim \xi \hbar/\tau_p$ that appears in Eq.~\eqref{j:ball:res:2} due to the scattering in the valence band. As a result, the RPGE without scattering in the valence band is smaller than the current in Eq.~\eqref{j:ball:compact} by a factor of
\[
\sim \frac{\hbar q^2\tau_p}{m_e}.
\]
For typical conditions this factor is small.

\section{Intersubband scattering-assisted photocurrent}

In noncentrosymmetric quantum well systems, the intersubband scattering probability of light $W^{sc}_{3\leftarrow 1}(\bm e, \bm k)$ contains, in general, an asymmetric part. As a result, there is a ballistic contribution  to the photocurrent:
\begin{equation}
\label{ballistic_intersubb}
\bm j  = {\frac{2e}{\mathcal S_0}}\sum_{\bm k,\bm k', {\bm q'}} [\bm v_{c3,\bm k} \tau_{p}(E_{c3,\bm k}) - \bm v_{c1,\bm k} \tau_{p}(E_{c1,\bm k})] W^{sc}_{3\leftarrow 1}(\bm e, \bm k) {f_0(E_{c1,\bm k})[1 - f_0(E_{c3,\bm k})]},
\end{equation}
where 
$\mathcal S_0$ is the normalization area.
Substitution of the scattering rate given by Eq.~(11) of the main text yields
\begin{multline}
\label{j:ball:res_intersubb}
j_\alpha =  {\frac{e}{\mathcal S_0}}  D_{\alpha\beta} \mathrm i [\bm e\times \bm e^*]_\beta \frac{4\pi}{\hbar^2} I |M_0|^4\\
\times  \sum_{\bm k,\bm q'} {\hbar^2 k^2\over 2} \frac{\tau_{p}(E_{c3,\bm k})/m_{c3} - \tau_{p}(E_{c1,\bm k})/m_{c1}}{(\hbar\omega - E_{c3,\bm k} + E_{v,\bm k})^2}
{f_0(E_{c1,\bm k})[1 - f_0(E_{c3,\bm k})]} \delta(\hbar\omega - \hbar\omega' - E_{c3,\bm k} + E_{c1,\bm k}).
\end{multline}

For the following calculations we perform the summation over $\bm q'$ by means of the energy conservation $\delta$-function:
\begin{equation}
\label{j:ball:res:1_intersubb}
j_\alpha =  {\frac{e}{\mathcal S_0}} D_{\alpha\beta} \mathrm i [\bm e\times \bm e^*]_\beta \frac{2}{\hbar} I |M_0|^2\gamma_r^{\rm QW} \sum_{\bm k} {\hbar^2 k^2\over 2}\frac{\tau_{p}(E_{c3,\bm k})/m_{c3} - \tau_{p}(E_{c1,\bm k})/m_{c1}}{(\hbar\omega - E_{c3,\bm k} + E_{v,\bm k})^2}
{f_0(E_{c1,\bm k})[1 - f_0(E_{c3,\bm k})]} ,
\end{equation}
where we introduced the rate of emission of the secondary photon (i.e., the recombination rate of the electron-hole pair in the vicinity of the fundamental band gap)
\[
\gamma_r^{\rm QW} = \frac{2\pi}{\hbar} |M_0|^2 \sum_{\bm q'} \delta(\hbar\omega - \hbar\omega' - E_{c3,\bm k} + E_{c1,\bm k}).
\]
Further calculations can be simplified as follows: 
we assume $\tau_{p}(E_{c3,\bm k}) \ll \tau_{p}(E_{c1,\bm k})$, ${f_0(E_{c3,\bm k}) \ll 1}$. Then we have
\begin{equation}
j_\alpha = - {\frac{e}{\mathcal S_0}} D_{\alpha\beta} \mathrm i [\bm e\times \bm e^*]_\beta \frac{2}{\hbar} I |M_0|^2\gamma_r^{\rm QW} \sum_{\bm k} \frac{E_{c1,\bm k}\tau_{p}(E_{c1,\bm k})}{(\hbar\omega - E_{c3,\bm k} + E_{v,\bm k})^2}
f_0(E_{c1,\bm k}) .
\end{equation}
For crude estimation we can assume $E_F, T \ll E_{g3} - \hbar\omega$, where $E_{g3}=E_g + \Delta_{31}$,
%and $T$ is the temperature expressed in the energy units,
 and obtain
\begin{equation}
j_\alpha \approx - e n_e \gamma_r^{\rm QW} D_{\alpha\beta} \mathrm i [\bm e\times \bm e^*]_\beta \frac{2\left<E\right> \tau_{\rm tr}}{\hbar} I  {|M_0|^2  \over (E_{g3} -\hbar\omega)^2}.
\end{equation}

More precisely, using
\begin{equation}
\sum_{\bm k}\ldots  = \mathcal S_0 g_{c1} \int_0^\infty \dd E_{c1} \ldots, \quad
E_{c3,\bm k} - E_{v,\bm k}= \nu_{31} E_{c1,\bm k} +E_{g3}, \quad \nu_{31}= m_{c1}\qty({1\over m_{c3}}+ {1\over m_h}), 
\end{equation}
where $g_{c1}$ is the density of states in the 1st conduction subband, we get
\begin{equation}
j_\alpha = - e D_{\alpha\beta} \mathrm i [\bm e\times \bm e^*]_\beta \frac{2}{\hbar} I |M_0|^2\gamma_r^{\rm QW} {\hbar\over 2\pi n_i\abs{V}^2} \int_0^\infty \dd E \frac{E}{(\hbar\omega - E_{g3} - \nu_{31}E)^2}
f_0(E) .
\end{equation}

At low temperatures we have:
\begin{equation}
\int_0^{E_\text{F}} \dd E \frac{E}{(\hbar\omega - E_{g3} - \nu_{31}E)^2}
= {1\over \nu_{31}^2} \qty(\ln{1+\varepsilon\over \varepsilon}-{1\over 1+\varepsilon}), \qquad \varepsilon = {E_{g3} -\hbar\omega\over \nu_{31}E_\text{F}},
\end{equation}
while the electron concentration is given by $n_e =2g_{c1}E_\text{F}$.
Therefore we get for Fermi statistics (when $\tau_\text{tr}=\tau_p(E_\text{F})$) Eqs.~(12),~(13) of the main text:
\begin{equation}
j_\alpha = - e n_e \gamma_r^{\rm QW} D_{\alpha\beta} \mathrm i [\bm e\times \bm e^*]_\beta   I { |M_0|^2\tau_\text{tr} \over \nu_{31}^2\hbar E_\text{F} }  \qty(\ln{1+\varepsilon\over \varepsilon}-{1\over 1+\varepsilon}).
\end{equation}

At Boltzmann statistics we have:
\begin{equation}
\int_0^\infty \dd E \frac{E f_0(E)}{(\hbar\omega - E_{g3} - \nu_{31}E)^2}
 = {\exp(\mu/T)\over \nu_{31}^2} \int_0^\infty \dd x \frac{x\exp(-x)}{(x+b)^2}
={\exp(\mu/T)\over \nu_{31}^2} \qty[\text{e}^b\text{Ei}(-b)(1+b)-1],
\end{equation}
where $\text{Ei}(x) = \int_{-x}^\infty t^{-1} \exp(-t) dt $ is the exponential integral,
\begin{equation}
b={E_{g3} -\hbar\omega\over \nu_{31}T},
\qquad
n_e= 2g_{c1} T \exp(\mu/T).
\end{equation}
As a result, we obtain
\begin{equation}
j_\alpha = - e n_e\gamma_r^{\rm QW} D_{\alpha\beta} \mathrm i [\bm e\times \bm e^*]_\beta   I   {|M_0|^2\tau_p(T) \over \nu_{31}^2\hbar T }  \qty[\text{e}^b\text{Ei}(-b)(1+b)-1].
\end{equation}

\section{Diagrammatic approach to RPGE}

It is instructive to consider the circular photocurrent generation in the diagrammatic approach.

\begin{figure}[h]
\includegraphics[width=0.99\textwidth]{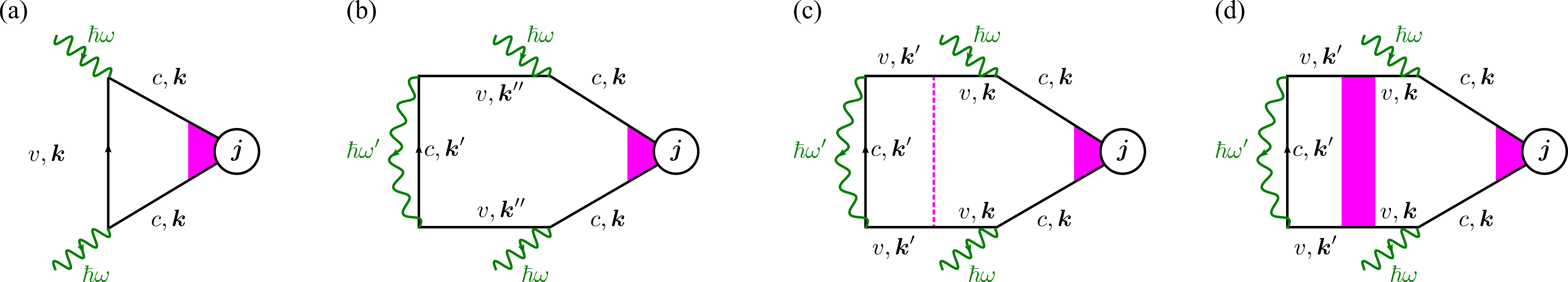}
\caption{Diagrams relevant for the photocurrent generation (in-scattering terms).}\label{fig:diag}
\end{figure}

Figure~\ref{fig:diag}(a) shows the diagram relevant for the CPGE at the light absorption. Taking into account that in the absorption region of the spectrum $\hbar\omega \geqslant E_g$ the energy conservation law can be fulfilled, Fig.~\ref{fig:diag}(a) can be immediately calculated and represents the conduction band contribution to the photocurrent, Eq.~\eqref{CPGE:0}. In the absence of absorption this diagram vanishes. 

\begin{figure}[h]
\includegraphics[width=0.2\textwidth]{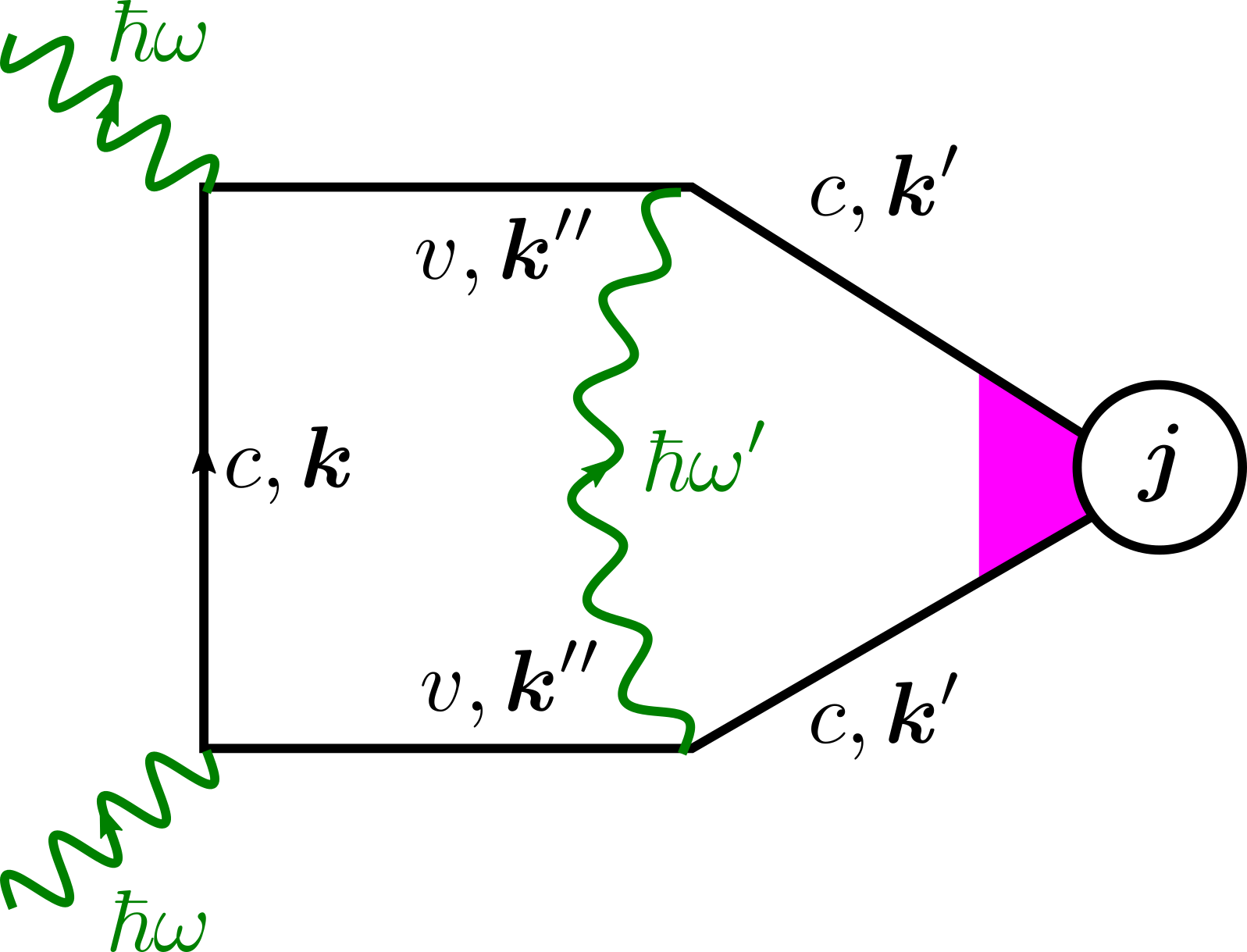}
\caption{Second diagram describing the RPGE with allowance for the photon wavevector (out-scattering term).}\label{fig:diag1}
\end{figure}

Diagrams~\ref{fig:diag}(b-d) describe the RPGE: These diagrams take into account the photon scattering, i.e., the emission of the secondary photon, wavy line marked as $\hbar\omega'$. 
The diagram Fig.~\ref{fig:diag}(b) and its counterpart, Fig.~\ref{fig:diag1}, describe the RPGE with allowance for the photon momentum. Note that $\bm k$, $\bm k''$ and $\bm k'$ are related by the momentum conservation law. Figure~\ref{fig:diag}(c) (and the diagram, analogous to that in Fig.~\ref{fig:diag1}) describe the photocurrent with allowance for the scattering in the valence band. The sum of all relevant diagrams with (0, 1, 2, \ldots) scattering is depicted in Fig.~\ref{fig:diag}(d).

\begin{figure}[h]
\includegraphics[width=0.45\textwidth]{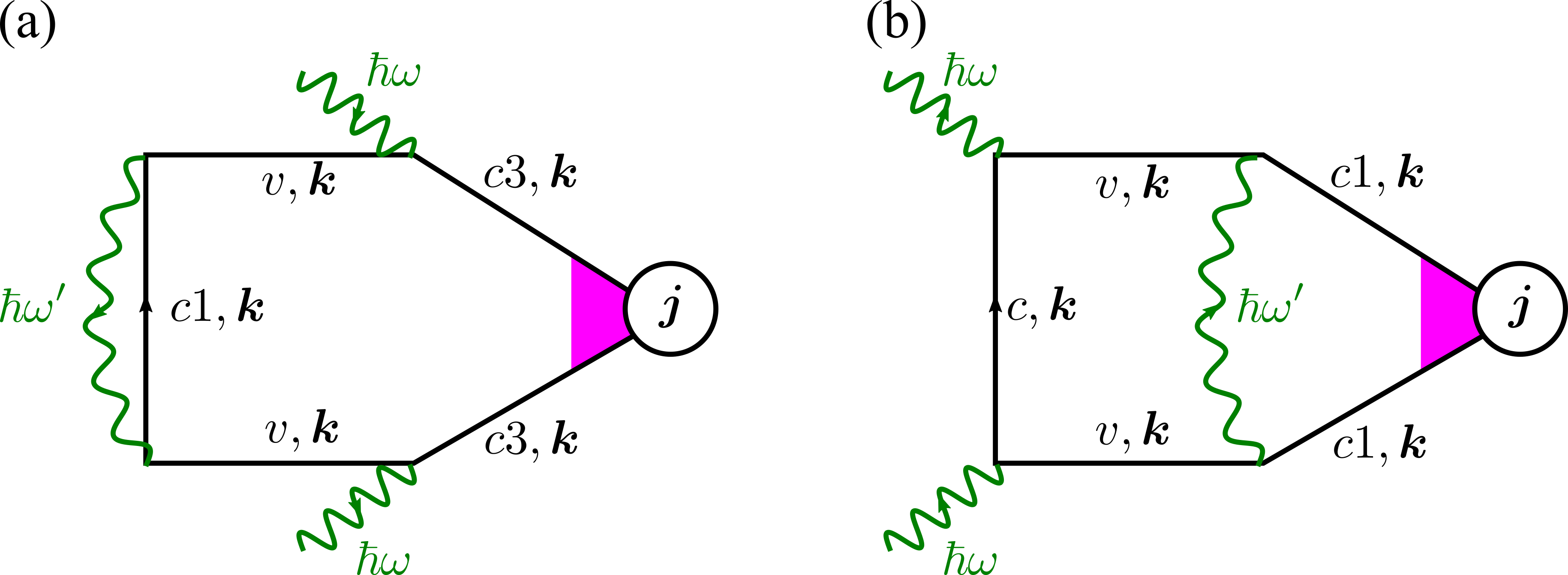}
\caption{Diagrams relevant for the RPGE at the intersubband scattering: (a) in-scattering and (b) out-scattering contributions.}\label{fig:diag2}
\end{figure}

Figure~\ref{fig:diag2} shows the diagrams relevant for the RPGE at the intersubband resonant scattering in quantum well structures. Here extra scattering in the valence band is not required, see Eq.~(11) of the main text.

\section{Photocurrents in the transparency region: general remarks}

It is instructive to make several comments about photocurrents for $\hbar\omega$ in the transparency region of the crystal regarding recent preprints~\cite{Onishi2022,shi2022berry}. We reiterate that in the absence of any real electronic transitions \emph{dc} current is forbidden. It is obvious from general reasons: If a \emph{dc} current is generated then this current results in a Joule heat in the sample or in the external circuit connected to the sample. It is forbidden by the energy conservation law in the absence of real transitions. On the microscopic level, the absence of the photocurrent follows from the arguments presented in Ref.~\cite{ISI:A1986E501100021}: In the absence of real transitions the only effect of the field is the renormalization of single particle dispersion $E_{c,\bm k} \to \tilde E_{c,\bm k}$. During the transient processes of energy and momentum relaxation the distribution function relaxes to an equilibrium function $f_0(\tilde  E_{c,\bm k})$ of the renormalized dispersion. The \emph{dc} current would be
\begin{equation}
\label{dc}
\bm j_{\rm dc} = 2\sum_{\bm k} \tilde v_{\bm k} f_0(\tilde  E_{c,\bm k}) =0,
\end{equation}
where $\tilde v_{\bm k} = \hbar^{-1} \partial \tilde E_{c,\bm k}/\partial \bm k$. Thus, real electronic transitions are crucial for the photocurrent generation.

In the case considered in our work the real transitions are induced by the light scattering (this point was mentionned in Ref.~\cite{Onishi2022}). In recent preprint~\cite{shi2022berry} the authors discussed the photocurrent in the optical gap of a metal related to the ``Berry curvature dipole'' and ``jerk'' effects. In this regard, two comments are due:

i) At linear polarization no current can be generated in the absence of absorption or scattering processes. It follows from the general relation~(1) of the main text:
\begin{equation}
\label{PGE}
j_\alpha = \gamma_{\alpha\beta} \mathrm i [\bm e\times \bm e^*]_\beta I + \chi_{\alpha\beta\mu} (e_\beta e_\mu^* + e_\mu e_\beta^*) I,
\end{equation}
which explicitly shows that the tensor $\chi$ responsible for the linear PGE is \emph{odd} at the time reversal. Thus, it contains odd numbers of dissipative constants.

ii)  The ``Berry curvature dipole'' contribution, at first glance, seems dissipationless: It appears at the circular polarization and does not vanish in the clean limit ($\Gamma\to 0$ in the terminology of the authors of Ref.~\cite{shi2022berry}). However, as demonstrated in Ref.~\cite{LG_EL_Spivak_PRB2020}, this contribution is related to the interference of \emph{real} electronic transition processes with the intermediate states in the same (conduction) band (Drude-like absorption) and with the intermediate states in the remote (valence) band. We also note that there is a side-jump contribution to the PGE which differs from the ``Berry curvature dipole'' contribution by a numerical factor only. Obviously, real transitions require dissipation, and as the authors of Ref.~\cite{shi2022berry} explicitly check, these processes do not violate basic thermodynamic principles. Note that the absence of scattering rates in the expressions for the ``Berry curvature dipole'' and side-jump contributions is due to the cancellation of the rates in product of the Drude transition probability and in the momentum scattering time entering the general expression for the photocurrent [cf. Eq.~(5) of the main text]:
\[
\bm j_{\rm dc} = e \bar{\mathbf{v}} \tau_p \dot N,
\] 
where for Drude-like transitions $\dot N \propto 1/\tau_p$ at $\omega\tau_p \gg 1$.

\bibliography{Raman}